\documentclass[sigconf]{acmart}
\renewcommand\footnotetextcopyrightpermission[1]{} 
\AtBeginDocument{%
  \providecommand\BibTeX{{%
    \normalfont B\kern-0.5em{\scshape i\kern-0.25em b}\kern-0.8em\TeX}}}

\begin{document}

\title{Multi-Mosaics: Corpus Summarizing and Exploration using multiple Concordance Mosaic Visualisations}


\author{Shane Sheehan}
\email{shane.sheehan@ed.ac.uk}

\affiliation{%
  \institution{University of Edinburgh}
  \country{United Kingdom}
}
\author{Saturnino Luz}
\email{s.luz@ed.ac.uk}
\affiliation{%
  \institution{University of Edinburgh}
  \country{United Kingdom}
}

\author{Masood Masoodian}
\email{masood.masoodian@aalto.fi}
\affiliation{%
  \institution{Aalto University}
  \country{Finland}
}

\renewcommand{\shortauthors}{}

\begin{abstract}
  Researchers working in areas such as lexicography, translation
  studies, and computational linguistics, use a combination of
  automated and semi-automated tools to analyze the content of text
  corpora. Keywords, named entities, and events are often extracted
  automatically as the first step in the analysis. Concordancing -- or
  the arranging of passages of a textual corpus in alphabetical order
  according to user-defined keywords -- is one of the oldest and still
  most widely used forms of text analysis. This paper describes
  Multi-Mosaics, a tool for corpus analysis using multiple implicitly
  linked Concordance Mosaic visualisations. Multi-Mosaics supports
  examining linguistic relationships within the context windows
  surrounding extracted keywords.
\end{abstract}


\begin{teaserfigure}
  \includegraphics[width=\textwidth]{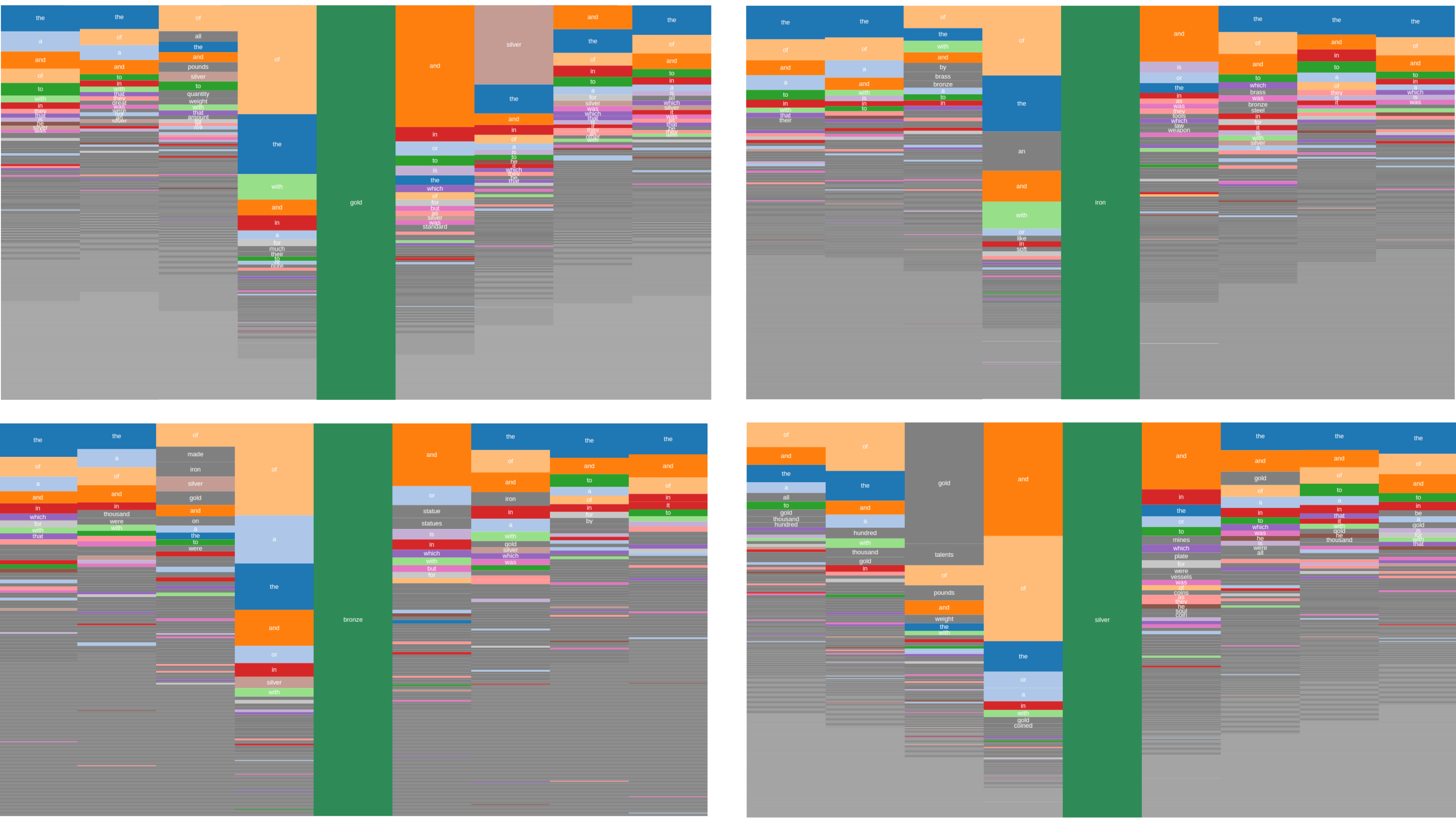}
  \caption{Top left Concordance Mosaic displaying the primary keyword \textit{gold}. Three secondary keywords (\textit{iron}, \textit{bronze} and \textit{silver}) are presented and colored to highlight the most frequent context words from the primary Mosaic.}
  \Description{}
  \label{fig:teaser}\vspace{3ex}
\end{teaserfigure}

\maketitle
\pagestyle{plain}
\section{Introduction}

In many academic fields, corpus analysis is central to the study of
texts.  Computational tools have long been used for lexicography,
corpus linguistics, and corpus-based translation studies
\cite{bib:Svartvik11d,bib:BernardiniKenny20c,bib:Bakerothers93c}, and
new methods motivated by such tools have been more broadly applied to
the study of policy in areas such as medicine
\cite{bib:ButsBakerLuzEngebretsen2021epistemologies} and politics
\cite{bib:Buts20crm}. One of the most popular techniques supported by
computation is the indexing, retrieval and display of
keyword-in-context. This technique dates back to at least the 1950's
with Luhn's work on concordance indexing \cite{bib:Luhn60k}.

Corpus analysis using concordance and collocation provides a data-driven approach for corpus analysis -- contrasting with more traditional scholarly work in these fields which requires close reading, researchers' prior knowledge, and theoretical frameworks to interpret texts. Data-driven corpus analysis techniques are heavily influenced by the work of John Sinclair and Michael Halliday \cite{bib:Leon07m,bib:Sinclair91CorpusConcordance}, in which one usually starts by obtaining an overview of the data and exploring a much larger volume of text than would be practical to do by close reading of the texts. Visualization tools can aid this process by providing
effective overviews and helping to identify patterns in the texts, as well as visual explanation of the analysis outputs \cite{bib:Tufte90}.

Concordance Mosaic \cite{bib:LuzSheehanAVI14} is a visualisation tool  which has been adopted by members of the corpus linguistic community for corpus analysis, and the presentation of their scholarly work \cite{bib:Baker20r,bib:Buts20crm} . The visualisation provides an overview of the context words within a window of the selected keyword. Quantitative information -- most often word frequency -- is encoded using tile height allowing the analyst to identify positional collocation patterns around a single keyword. 

This paper presents Multi-Mosaics, as an extension of Concordance Mosaic, to enable comparison of collocation patterns for multiple keywords. The design of Multi-Mosaics has been informed by observations, and requirements gathered using Concordance Mosaic. This paper also presents a user study where the Concordance Mosaic is compared to a traditional keyword-in-context textual display.

\section{Related Work}
Keyword visualisations often take the form of networks  or clusters of keywords\cite{bib:VisKeywords,bib:Li2016,CHOI2014}. While these visualisations are useful for identifying similar or connected keywords in a corpus, they do not provide any insight into the collocations of the keywords, and as such, they are not directly usefully for comparing the contexts in which these keywords appear.

\textit{Corpus Clouds} \cite{bib:Culy2011} is a frequency-focused
corpus analysis tool. A word cloud, based on the tag cloud
visualisation \cite{bib:ViegasWattenberg08tt}, is used to encode the frequencies of all words returned by a corpus query. For quantitative tasks involving frequency estimation or comparison, the use of font size to encode value in cloud-based visualisations is a limitation  \cite{bib:WordCloudEVAL}. Also, since positional collocations are not encoded in \textit{Corpus Clouds}, the visualisation is of an entire context window.

\textit{TagSpheres} \cite{bib:TagSpheres} are word cloud based visualisation where keyword co-occurrences are encoded using an integral combination of color and radial position from the central keyword. The cloud layout places the same word from different positions close together to help identify strong collocation patterns. However, the linear structure of the text is removed making it difficult to identify multi-word collocation patterns.

Tree-based visualisations of keyword-in-context have therefore been proposed as a technique for encoding quantitative information while maintaining the linear structure of the text \cite{bib:Culy2010,bib:LuzSheehanAVI14,bib:WattenbergViegas08}. In practice, however, displaying a large number of concordance lines, where font size is used to encode represent positional frequency, requires a trade off between frequency estimation and readability. The variable length of words also makes encoding frequency using font size challenging -- since area is not as perceptually efficient as length  for visualising quantitative information.

\section{Multi-Mosaics}

\begin{figure*}[h]
  \centering
  \includegraphics[width=.7\linewidth]{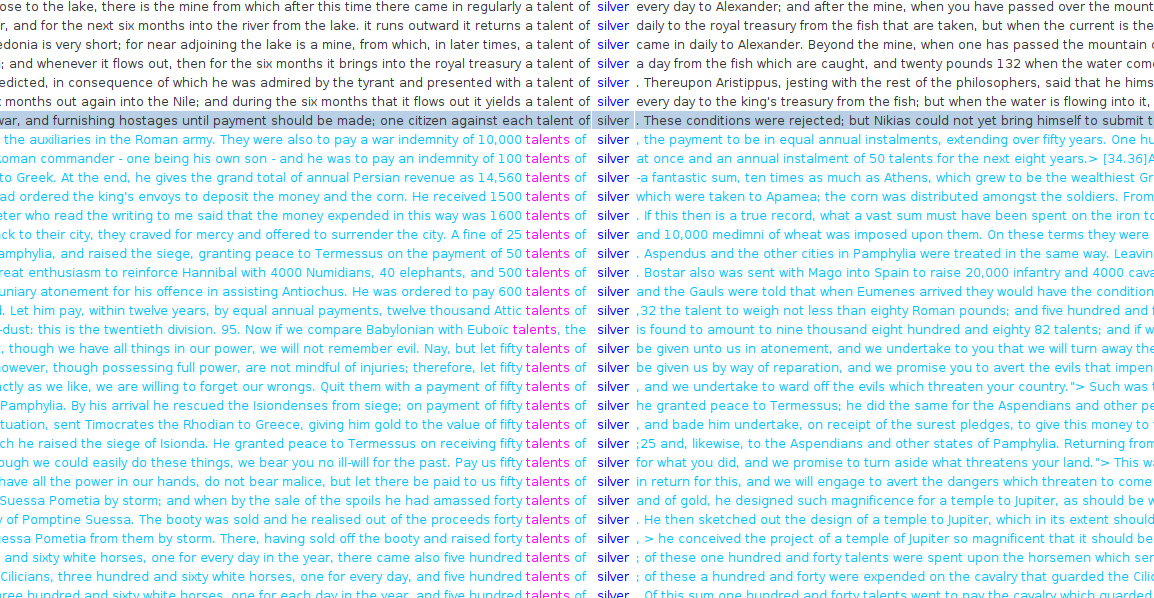}
  \caption{Textual keyword-in-context linked to Multi-Mosaics. The Mosaic for the keyword \textit{silver} is selected and the word \textit{talents} at position keyword minus 2 was chosen using the Mosaic.}
  \Description{concordance}
   \label{fig:concordance}
\end{figure*}

Multi-Mosaics was designed in collaboration with language scholars who work predominantly using corpus analysis techniques. During this co-design process two of the researchers were asked to compile a list 20 questions which they would like to be able to answer about a corpus. They were asked to list them in order of importance, and after the lists were compiled they were used to discuss requirements for a potential visualisation prototype.
This technique is a common means of requirements elicitation in human-computer interaction \cite{bib:Requirments}.

This process revealed a need for a tool which facilitated quantitative analysis of the words surrounding a keyword \cite{luz2020methods}. Concordance Mosaic was developed to address this need, where each mosaic visualization displays quantitative information -- such as word frequency or collocation strength -- in tiles representing words at positions relative to a keyword. The height of each word tile encodes the quantity, and the horizontal position of the columns represents  word position relative to the keyword. In Figure~\ref{fig:teaser} (top left) we see that the most frequent word occurring directly to  the left of the keyword \textit{gold} is the word \textit{of}.

The language scholars also mentioned the need to investigate multiple keywords for similar patterns or to investigate lists of suggested keywords for a corpus. The following suggests that investigating collocation patterns of multiple words simultaneously would be useful ``\textit{...if the keyword is a label used to describe a particular kind of political agent, we might be interested to look at what collective nouns are used to group and characterize these political agents (e.g. a mob of citizens, a tribe of politicians: LEFT +2).} ". Similarly it was suggested that automatically extracting related keywords would be useful to expand the analysis to  include the collocations of these keywords ``\textit{Are there other related keywords we might study in order to expand our investigation? Can the software suggest keywords that are important to these texts but which we might not otherwise have thought of?}". Using these observations the concept of multiple mosaics displayed together with linked  contexts was presented to the corpus linguists and the design was refined to produce Multi-Mosaics. 

\subsection{Implementation}

Multi-Mosaics\footnote{The code and example file format can be found at \url{https://github.com/sfermoy/MultiMosaic}} is implemented as a single-page web application using the \emph{D3.js} framework \cite{bib:Bostock2011}. The interface displays a grid of Concordance Mosaics, generated from JSON files containing keyword-in-context data, and a single textual keyword in context window. 

There is no theoretical limit to the number of Concordance Mosaic visualisations which can be displayed in Multi-Mosaics, as the interface is scrollable. 
Each mosaic displays a single keyword and four word positions to either side of the keyword. The Mosaic displayed in the top left position of Multi-Mosaics is considered the primary Mosaic, and all others are secondary Mosaics. Right clicking on a secondary Mosaic makes it the primary, and updates Multi-Mosaics. In Figure \ref{fig:teaser} the primary Mosaic for the keyword \textit{gold} is selected.

Each mosaic is colored according to the 20 most frequent words in the 4-word context window of the primary mosaic. This colouring allows the corpus analyst to investigate each of the secondary Mosaics contexts in relation to the high frequency context words from the primary Mosaic. Words not in the 20 most frequent context words of the primary Mosaic are colored grey. When looking at the contexts of secondary Mosaics, grey word tiles with high positional frequencies are also likely to be of interest, since they represent words which have high secondary keyword positional frequency, and low primary keyword positional frequency. In Figure \ref{fig:teaser}, looking at the secondary mosaic for the keyword \textit{silver} (bottom right), in relation to the primary Mosaic for the keyword  \textit{gold}, we can identify interesting collocation patterns. For instance, we  can see that the word \textit{gold} is a frequent collocate at position $keyword -2$, and we can find \textit{silver} in position $keyword +2$ for the Mosaic of \textit{gold}. The second most frequent word at position $keyword -2$ is the word \textit{talents} -- this word is is not in the 20 most frequent words in the context of the primary keyword \textit{gold} so it is coloured grey.

In addition to the grid of Mosaics, a single textual keyword-in-context concordance window is also available, as shown in Figure \ref{fig:concordance}. Initially this window displays the concordance lines from the primary keyword Mosaic. However, left clicking on any tile, in any of the Mosaics, switches the textual keyword-in-context window to display the concordance lines for the clicked keyword. The keyword from the selected Mosaic is displayed centrally and colored Blue -- in Figure \ref{fig:concordance} the Mosaic for the keyword \textit{silver} was clicked. The tile which was clicked represented the word \textit{talents} at  the position $keyword -2$, the concordance window is sorted alphabetically at this position and the lines containing the clicked word at the chosen position are highlighted in Cyan, the work which was clicked is coloured Pink. This enables quick investigation of the lines from the corpus which form the identified collocation pattern, in an overview and detail-on-demand interaction \cite{bib:Shneiderman1996}. In addition the Concordance Mosaic which was clicked will have all words connection to the keyword highlighted in white, this helps with identifying multi-word patterns of collocation.


\section{User study}

This study was designed to compare the performance of the two visualisation tools, the Concordance Mosaic and a textual Keyword-in-Context interface (KWIC). A third option was also tested, in which both the Concordance Mosaic and KWIC were available side-by-side. The evaluation was performed on concordance analysis tasks for which the Concordance Mosaic was designed. These tasks have been identified from analysis of the corpus methodology \cite{luz2020methods}, described by John Sinclair \cite{bib:Sinclair03Reading}. 

An initial heuristic evaluation and a pilot study were used to refine the visualization tools and test their usability prior to the main user study. During this study we found that tasks requiring analysis of multiple context words or positions were difficult for non-expert users to understand. Based on this we limited the evaluation to five simple quantitative analysis tasks, only one of which required looking at multiple positions.

Each participant attempted to answer five questions using each of the 3 visualisation options -- the order in which the options were presented was 
randomised and balanced across participants for every possible combination of option orderings. Similarly, for each visualisation option a different keyword was required per question, and the keyword per option was balanced. 

Question 1 was, for example, \textit{``For the keyword KEYWORD, what is the most frequent word at position keyword - 1?"}.
The three possible keywords chosen for this question were \textit{Wealthy}, \textit{Daylight} and \textit{Massive}. These were chosen to ensure a consistent number of concordance lines were returned from the corpus used for this study, and that the difficulty level of the questions was consistent. For question one these keywords all returned a 
concordance with approximately 300 concordance lines -- the most frequent word at position $keyword -1$ occurs with a frequency of between 26--27\%, and the second most frequent word 
at position $keyword -1$ occurs with a frequency of 20--22\%. Question 2 was the same as the first one, but the frequencies of the most common and second most common words at position $keyword -1$ were approximately 40\% and between 5--10\%, respectively -- thus making this task more difficult.

The remaining three task questions were,
Question 3: \textit{``For the keyword KEYWORD, what is the most frequent descriptive adjective at position $keyword -1$?"},
Question 4: \textit{``For the keyword KEYWORD, focusing only on concordances that contain the word CONTEXT-WORD at position $keyword -1$, what word is most frequent at position $keyword -2$?"},
Question 5: \textit{``Estimate which word has the highest collocation strength at position $keyword -1$"}. For the last question, collocation strength was described as high frequency at a position but low frequency in the corpus. 

The participants were shown how to access the corpus frequency lists prior to the study. It was possible to switch the Mosaic to collocation strength mode, in which the tiles were scaled according to this metric.

The null hypothesis in this study was that: \textit{there is no significant 
difference in performance between the three visualisation options, on concordance analysis tasks}. Performance was measured using the speed and accuracy with which participants completed tasks.

For this study we recruited 36 participants from the student population through our online university noticeboard and a mailing list. Since the study evaluated performance on quantitative tasks, we decided that previous experience with concordance tools or corpus analysis would not be a prerequisite for participation.

\subsection{Results}
Table \ref{tab:Anova} shows the results of an ANOVA for the dependant variable \textit{t}, time to complete 
a task (in seconds), with respect to the categorical variables: the question being answered (q), the visualisation being used (i), the participants assigned visualisation option ordering (iOrder), the participants assigned keyword set ordering (qOrder) and a binary variable 
representing a correct or incorrect answer (isCorrect). The results of the ANOVA where a significant difference ($p < .05$) was found are also shown.

\begin{table}[h]
  \begin{center}
  \footnotesize
  \caption{ANOVA results for the dependant variable time \textit{t}, where p $<$ .05 }
  \label{tab:Anova}
		\begin{tabular}{ l  c  r }
  			\hline               
          
  			Independent Variable & F Value & P Value \\
  			\hline
  			q  & 26.9388 & $< 2.2e^{-16}$  \\
			i & 135.5089 &$< 2.2e^{-16}$  \\
  			isCorrect & 4.3170 & 0.038894 \\ 
  			q:i &8.0946 & $1.428e^{-9}$ \\
  			i:iOrder  & 2.8620 & 0.002261 \\
  			i:qOrder  & 3.5232 & 0.008239 \\ 
  			q:isCorrect & 2.4258 & 0.048970 \\
 		 \hline  
		\end{tabular}
	\end{center}
\end{table}

Since the main effects q, i and isCorrect all feature in significant interactions, we have focused our post-hoc analysis on these interactions instead of the main effects. We conducted Tukey's post-hoc tests (HSD) to analyse the different groupings of each interaction effect, again using p $<$ .05 to test for significance.

The result of the HSD test for the i and qOrder interaction (i:qOrder) showed a significant difference between two groupings. In this case, the data set was split into 9 groups by the combinations of the 3 visualisation options and the 3 circularly shifted keyword set orderings. The HSD groupings simply combined these groups into data points where the KWIC option was being used, and a grouping of all data points where either the Mosaics or combined visualisations were being used.  This indicates that the interaction can be interpreted as i, and that qOrder can be safely ignored, as it does not feature in any other significant interactions or as a main effect. This result shows, as expected, that our choice of keywords has not had a major effect on time to complete within each question.

The mean response times of the i:qOrder groups in which the KWIC option was used were all greater than 67 seconds, while the remaining groups containing the Mosaic and combined option all had mean response times under 30 seconds. This is evidence of visualisation options having a large effect on response time.

The discovery of an interaction between q and i (q:i) is of great interest since our null hypothesis states: there are no significant differences between the interfaces on a per question basis.
Analysing the groups created by splitting the data by visualisation options and questions, the Tukey HSD test found a number of significant groupings. For each question there is a significant difference in response time between the KWIC and both the Mosaics and combined options. This is enough evidence to reject our null hypothesis for each question.

\section{Conclusions}
In this paper, we introduced Multi-Mosaics, a visualisation tool consisting of multiple linked Concordance Mosaics and a textual keyword-in-context window. Multi-Mosaics is designed to support the analysis of collocation patterns for multiple keywords simultaneously. By encoding the most frequent collocations from the primary keyword Mosaic onto all secondary Mosaics similar and differing patterns of collocation can be identified in the secondary keywords. In addition, linking the Mosaics to a textual Keyword-in-Context concordance display allows for quick investigation of the lines of text making up the identified collocation pattern. 
Our user study evaluating the effectiveness of the visualisation tool shows that Concordance Mosaics perform better than a textual keyword-in-context tool on a selection of quantitative corpus analysis tasks.

\begin{acks}

\end{acks}

\bibliographystyle{ACM-Reference-Format}
\bibliography{multimosaic}

\end{document}